\documentstyle[12pt]{article}
\begin{document}
\noindent
{\large\bf Regional Centres for Space Science and Technology Education
(Affiliated to the United Nations)}\\
\begin{center} 
{\bf Hans J. Haubold}\\
Programme on Space Applications\\
Office for Outer Space Affairs\\
United Nations\\
Vienna International Centre\\
P.O. Box 500\\
A-1400 Vienna, Austria\\
Email: haubold@kph.tuwien.ac.at
\end{center}
\noindent
Abstract\par
\medskip
\noindent
Education is a prerequisite to master the challenges of space science and technology. Efforts to understand and control space science and technology are necessarily intertwined with social expressions in the cultures where science and technology is carried out (Pyenson [1]). The United Nations is leading an effort to establish regional Centres for Space Science and Technology Education in major regions on Earth. The status of the establishment of such institutions in Asia and the Pacific, Africa, Latin America and the Caribbean, Western Asia, and Eastern Europe is briefly described in this article.\par
\medskip
\noindent
{\bf 1. United Nations Programme on Space Applications}\par
\smallskip
\noindent
The United Nations Programme on Space Applications was established in 1971 on the recommendation of the first United Nations Conference on the Exploration and Peaceful Uses of Outer Space (UNISPACE I)[2] and the Programme was expanded and its mandate broadened in UNISPACE II (1982) [3] and the recently concluded UNISPACE III [4] Conferences. Fulfilling one element of the Programme's mandate, more than 150 workshops with approximately 8000 participants have been organized since its establishment. Following the need of developing countries and taking into account the space-related agenda of the Programme, the majority of workshops focussed on core disciplines: remote sensing and geographic information system, satellite communications and geo-positioning system, satellite meteorology and global climate, and space and atmospheric sciences  [5]. Despite the success of these workshops in the initiation of regional and international cooperation and the development of space science and technology, particularly for the benefit of developing countries, in the 1980's the limitations of short-term activities were recognized and called for the need of building long-term regional capacity in space science and technology and its applications [6]. Subsequently, in 1988, under the auspices of the Programme, a project to establish centres for space science and technology education at the regional level was initiated [7]. A unique element of this project was that the Centres were envisaged to be established in developing countries for the benefit
of regional cooperation, particularly between the developing countries.\par
\medskip
\noindent
{\bf 2. United Nations General Assembly Resolutions}\par
\smallskip
\noindent
The General Assembly of the United Nations, in its resolution 45/72 of 11 December 1990, endorsed the recommendation of the Working Group of the Whole of the Scientific and Technical Subcommittee, as approved by the Committee on the Peaceful Uses of Outer Space (COPUOS) [8], that: ``... the United Nations should lead, with the active support of its specialized agencies and other international organizations, an international effort to establish regional centres for space science and technology education in existing national/regional educational institutions in the developing countries'' [9]. \par
\smallskip
\noindent
Subsequently, the General Assembly, in its resolution 50/27 of 6 December 1995, also
endorsed the recommendation of COPUOS that ``these centres be established on the basis of affiliation to the United Nations as early as possible and that such affiliation would provide the centres with the necessary recognition and would strengthen the possibilities of attracting donors and of establishing academic relationships with national and international space-related institutions'' [10]. \par
\medskip
\noindent
{\bf 3. Status of Establishing and Operating the Regional Centres}\par
\smallskip
\noindent
At the occasion of the UNISPACE III Conference (19-30 July 1999, Vienna, Austria), the status of the operation and establishment of the regional Centres was reviewed as part of the intergovernmental meetings and the technical forum of this Conference [4].\par
\smallskip
\noindent
Since its inauguration in India in 1995, the regional Centre for Space Science and Technology Education in Asia and the Pacific has successfully conducted four post-graduate courses on remote sensing and geographic information system; two courses on satellite communications; and a course each on the following topics: satellite meteorology and global climate; and  space science. Each of the courses was inaugurated through a research level workshop on the respective topic supported through regular activities of the United Nations Programme on Space Applications. Upon completion of the nine-month course in each activity, the scholars have carried out a one-year applications/research project in their
home countries. In agreement with resolution 45/72, this Centre takes advantage of the intellectual resources and facilities of three renowned space-related institutions: (i) the Indian Institute of Remote Sensing, Dehradun, (ii) the Space Applications Centre, Ahmedabad, and (iii) the Physical Research Laboratory, Ahmedabad [15].\par
\smallskip
\noindent
The regional Centre for Space Science and Technology - in French Language - in Africa was inaugurated on 24 October 1998 in Casablanca, Morocco, and is located at the Ecole Mohammadia d'Ingenieurs in Rabat.\par
\smallskip
\noindent
The regional Centre for Space Science and Technology Education - in English Language - in Africa was inaugurated on 24 November 1998 in Abuja, Nigeria, and is located at Obafemi Awolowo University in Ile-Ife [16].\par
\smallskip
\noindent
The inauguration of the regional Centre for Space Science and Technology Education in Latin America and the Caribbean is expected to occur in 2000 in Brazil and Mexico. In preparation for the operation of the campus of the Centre in Brazil, the Instituto Nacional de Pesquisas Espaciais (INPE) is already very active in carrying out a number of workshops for the benefit of States in the region.\par
\smallskip
\noindent
An evaluation mission to Jordan and the Syrian Arab Republic was conducted in 1998. The reports of the mission have been finalized in consultation with the Governments of Jordan and the Syrian Arab Republic, with a view to selecting a host country for a regional Centre in Western Asia, which is expected to occur shortly after the UNISPACE III Conference.\par
\smallskip
\noindent
In 1995, the Network of Space Science and Technology Education and Research Institutions for States of Central-Eastern and South-Eastern Europe was established [11]. A technical study mission to Bulgaria, Greece, Hungary, Poland, Romania, Slovakia, and Turkey was carried out in 1998. The mission undertook a technical study and provided an informative report that will be used in determining, in each country visited, an agreed framework for the operation of such a Network. Each country designated space science and technology related core and associated institutions, all of them with a long and successful history
in research and applications of space science and technology, which are being part of this Network.\par
\clearpage
\noindent
{\bf 4. Governing Boards and Advisory Committees of the Centres}\par
\smallskip
\noindent
Each Centre shall aspire to be a highly reputable regional institution, which, as the needs arise, and as directed by the Centre's Governing Board, may grow into a network of specialized and internationally acclaimed affiliate nodes. 
Because resolution 45/72 specifically limits the role of the United Nations to ``lead, ..., an international effort to establish regional centres'', it is apparent that once a Centre is inaugurated, its Governing Board will assume all decision-making and policy-formulating responsibilities for the Centre. The Governing Board is the overall policy making body of each Centre and consists of member States (within the region where the Centre is located), that have agreed, through their endorsement of the Centre's agreement, to the goals and objectives of the Centre. The agreement of the Centre calls for the establishment of an
Advisory Committee that provides advise to the Governing Board on all scientific and technical matters, particularly on the Centre's education curricula,  and consists of experts in the field of space science and technology [12]. The United Nations serves the Centre and its Governing Board and Advisory Committee in an advisory capacity. Governing Boards were established for the Centres in Asia and the Pacific and Africa. To date the Advisory Committee has been set up for the Centre in Asia and the Pacific.\par
\medskip
\noindent
{\bf 5. Next Steps to Be Taken}\par
\smallskip
\noindent
During the deliberations of the UNISPACE III Conference, meetings were held and
presentations were delivered to chart the course for future measures to continue
furthering the regional Centres. In a meeting between representatives of the Centres in Asia and the Pacific, Africa, and Latin America and the Caribbean, the opinion was emphasized, that as a follow-up of the Conference, closer and lively cooperation between the regional Centres needs to be established already at this point of time. Particularly, the rich experience gained in the successful operation of the Centre in Asia and the Pacific as centre of excellence shall be made available to the Centres in all other regions. It was further felt that all Centres, through the support of the United Nations Office for Outer Space Affairs and its Programme on Space Applications, should urgently
establish cooperation with international organizations and institutions (among them COSPAR, IAU, ICTP, ISPRS, ISU, TWAS), specialized agencies of the United Nations system (among them FAO, IAEA, UNESCO, UNU, WHO, WMO), and the Economic and Social Commissions of the respective region. The International Astronomical Union (IAU) has undetaken first steps in this direction [13]. The strong participation of developing countries in the technical forum activities of UNISPACE III also brought to the attention of the Office for Outer Space Affairs that the Centre's education curricula [12] may have to be supplemented with non-core discipline elements focussing on space biology/medicine, devising small  satellite projects, microgravity, and other space-related topics. \par
\medskip
\noindent
{\bf 6. UN/ESA Workshops on Basic Space Science}\par
\smallskip
\noindent
The establishment of the regional Centres is the sole project of the Programme on Space Applications leading to ``institutionalization'' in the field of space science and technology. The operation of the Centres can be supported by the Programme in organizing some of its regular activities in close cooperation with the Centres. In this connection it shall be recalled that it was India in 1991, hosting the first United Nations/European Space Agency Workshop on Basic Space Science for the benefit of Asia and the Pacific at ISRO in Bangalore, that inaugurated a series of worldwide workshops. Since then such workshops were organized in Latin America and the Caribbean (Costa Rica and Colombia 1992,
Honduras 1997), Africa (Nigeria 1993), Western Asia (Egypt 1994, Jordan 1999), Europe (Germany 1996, France 2000), and again in Asia and the Pacific (Sri Lanka 1995) [14]. This series of workshops led to the establishment of several education and research oriented astronomical telescope facilities with a view to link them to the respective regional Centres in the future. Already such a series of workshops, organized in the field of space science and technology, can lead to an appreciable expansion of cooperation between countries of a region and its regional Centre.\par
\medskip
\noindent
{\bf 7. Contact Adresses for More Details on the Regional Centres and Their Education
Programmes}\par
\noindent
{\it Asia and the Pacific Region}\\
Prof. B. Deekshatulu\\
Centre for Space Science and Technology Education in Asia and the Pacific
Indian Institute of Remote Sensing Campus\\
4 Kalidas Road\\
Dehra Dun - 248 001\\
India\\
Tel.: (+91)-135-740-737\\
Fax : (+91)-135-740-785\\\nopagebreak
Email: deekshatulu@hotmail.com\par
\medskip
\noindent
{\it Africa Region}\\
Prof. E.E. Balogun\\
Centre for Space Science and Technology Education - in English Language - in Africa
Department of Physics\\
Obafemi Awolowo University\\
Ile-Ife\\
Nigeria\\
Tel.: (234)-36-230-454\\
Fax : (234)-36-233-973\\
Email: ebalogun@oauife.edu.ng\par
\medskip
\noindent
{\it Africa Region}\\
Prof. A. Touzani\\
Centre Regional Africain des Sciences et Technologie de l'Espace Langue Francaise
Sis a l'Ecole Mohammadia d'Ingenieurs\\
Avenue Ibn Sina\\
B.P. 765, Agdal\\
Rabat\\
Maroc\\
Tel.: (212)-7-681824\\
Fax : (212)-7-681826\\
Email: craste@emi.ac.ma\par
\medskip
\noindent
{\it Latin America and the Caribbean Region}\\
Dr. T.M. Sausen\\
Instituto Nacional de Pesquisas Espaciais\\
Divisao de Sensoriamento Remoto\\
Av. dos Astronautas, 1758\\
Cx.P. 515\\
CEP 12201-970 Sao Jose dos Campos, SP\\
Brazil\\
Tel.: (+55)-12-325-6862\\
Fax : (+55)-12-325-6870\\
Email: tania@ltid.inpe.br\par
\medskip
\noindent
{\it Western Asia Region}\\
To be made available shortly\par
\bigskip
\noindent
{\bf Acknowledgements}\par
\smallskip
\noindent
The cooperation with Dr. W. Steinborn (German Space Agency, DLR) during the evaluation mission through Africa, Drs. G. Arrigo and B. Negri (Italian Space Agency, ASI) during the technical study mission through Central-Eastern and South-Eastern Europe, and Prof. F.R. Querci (French Space Agency, CNES) during the evaluation mission through the Middle East, is greatly acknowledged.\par
\bigskip
\noindent
{\bf References}\par
\medskip
\noindent
Note: The author is writing in his personal capacity and the views expressed in this paper are those of the author and not necessarily of the United Nations.\par
\smallskip
\noindent
[1] L. Pyenson and S. Sheets-Pyenson, Servants of Nature: A History of Scientific Institutions, Enterprises, and Sensibilities, W.W. Norton \& Company, New York, 1999, pp. XV+496.\par
\smallskip
\noindent
[2] United Nations Conference on the Exploration and Peaceful Uses of Outer Space, Vienna, 14-27 August 1968, United Nations, New York, 1968, Document E.68.I.11, pp. 59.\par
\smallskip
\noindent
[3] United Nations Conference on the Exploration and Peaceful Uses of Outer Space, Vienna, Austria, 9-21 August 1982, United Nations, New York, 1982, Document A/CONF.101/10, pp. 167; R. Chipman (Ed.), The World in Space: A Survey of Space Activities and Issues Prepared for UNISPACE 82, Prentice-Hall, 1982, pp. 689.\par
\smallskip
\noindent
[4] United Nations Conference on the Exploration and Peaceful Uses of Outer Space, Vienna, Austria, 19-30 July 1999, United Nations, Vienna, 1999, Document A/CONF.184/6; http://www.un.or.at/OOSA/.\par
\smallskip
\noindent
[5] Space for Development: The United Nations Programme on Space Applications, United Nations, Vienna, 1999, Document V.98-57085, pp. 23; http://www.un.or.at/OOSA/.\par
\smallskip
\noindent
[6] Report on the UN Workshop on Space Science and Technology and its Applications within the Framework of Educational Systems, 4-8 November 1985, Ahmedabad, India, Document A/AC.105/365, (27 December 1985) pp. 24; Report of the UN Meeting of Experts on Space Science and Technology and its Applications within the Framework of Educational Systems, 13-17 October 1986, Mexico, D.F., Document A/AC.105/378, (23 December 1986) pp. 25; Report on the UN Meeting of Experts on Space Science and Technology and its Applications within the Framework of Educational Systems, 27 April - 1 May 1987, Lagos, Nigeria,
Document A/AC.105/390, (18 November 1987) pp. 23; Report on the UN International Meeting of Experts on the Development of Remote-Sensing Skills and Knowledge, 26-30 June 1989, Dundee, United Kingdom, (3 January 1990) pp. 21.\par
\smallskip
\noindent
[7] Centre for Space Science and Technology Education, United Nations, New York, 1990, Documents SAP/90/001 to 003, pp. 24; Centres for Space science and Technology Education: A Progress Report, Document A/AC.105/498, (12 March 1990) pp. 28; Centres for Space Science and Technology Education: Updated Project Document, Document A/AC.105/534, (7 January 1993) pp. 56; Regional Centres for Space Science and Technology Education (Affiliated to the United Nations), Document A/AC.105/703, (16 June 1998) pp. 12.\par
\smallskip
\noindent
[8] M. Benkoe and K.-U. Schrogl, International Space Law in the Making: Current Issues in the UN Committee on the Peaceful Uses of Outer Space, Editions Frontiers, Gif-sur-Yvette, 1993, pp. XXIII+275.\par
\smallskip
\noindent
[9] Report of the Committee on the Peaceful Uses of Outer Space, General Assembly, Official Records: Forty-Fifth Session, Supplement No. 20 (A/45/20), United Nations, New York, 1990; Report of the Scientific and Technical Sub-Committee on the Work of its Twenty-Seventh Session, Document A/AC.105/456, (12 March 1990) pp. 37.\par
\smallskip
\noindent
[10] Report of the Committee on the Peaceful Uses of Outer Space, General Assembly, Official Records: Fiftieth Session, Supplement No. 20 (A/50/20), United Nations, New York, 1995.\par
\smallskip
\noindent
[11] M.-I. Piso, in Proceedings of the UNISPACE III Regional Preparatory Conference for Eastern Europe, Bucharest, Romania, 25-29 January 1999, published by the Romanian Space Agency under the auspices of the United Nations Office for Outer Space Affairs, Bucharest, Romania, 1999, pp. 185-198.\par
\smallskip
\noindent
[12] Centres for Space Science and Technology Education: Education Curricula, United Nations, Vienna, 1996, Document A/Ac.105/649, 23 pp.; Report on the UN/ESA/COSPAR Workshop on Data Analysis Techniques, 10-14 November 1997, Sao Jose dos Campos, Brazil, (19 December 1997) pp. 10.\par
\smallskip
\noindent
[13] Conclusions and Proposals of the IAU/COSPAR/UN Special Workshop on Education in Astronomy and Basic Space Science, 20-23 July 1999, UNISPACE III Conference, Document A/CONF.184/C.1/L.8, (23 July 1999) pp. 2; see also [4].\par
\smallskip
\noindent
[14] H.J. Haubold and W. Wamsteker, Space Technology 18(1998)No. 4-6, pp. 149-156; H.J. Haubold, Journal of Astronomical History and Heritage 1(1998) No. 2, pp. 105-121; http://www.seas.columbia.edu/$\sim$ah297/un-esa/.\par
\smallskip
\noindent
[15] Centre for Space Science and Technology Education (Affiliated to the United Nations) in Asia and the Pacific, Brochure issued by the Centre, Dehra Dun, India, 1995, pp. 6.\par
\smallskip
\noindent
[16] Centre for Space Science and Technology Education (Affiliated to the United Nations) in Africa, Brochure issued by the Centre, Ile-Ife, Nigeria, 1998, pp. 14.
\end{document}